\documentclass[twocolumn,showpacs,preprintnumbers,amsmath,amssymb]{revtex4}

\usepackage{graphicx}
\usepackage{dcolumn}
\usepackage{bm}

\begin{document}


\title{Spontaneous emission of homogeneous broadening molecules
on a micro-droplet's surface and local-field correction\\}

\author{Yun-Jie Xia$^{1,2,}$}
\email{yjxia@qfnu.edu.cn}
\author{Guang-Can Guo$^{1,}$}
\email{gcguo@ustc.edu.cn}
\affiliation
{
\centerline{ $^{1}$Key Laboratory of Quantum Information,
University of
Science and Technology of China,Chinese Academy of Sciences,\\
            }
\centerline{
Hefei,Anhui,China,230026.\\
           }
\centerline{
$^{2}$Department of Physics,Qufu Normal University,Qufu,Shandong,China,273165.\\
           }
}
\author{P. T. Leung}
\email{ptleung@phy.cuhk.edu.hk} \affiliation{
\centerline{Department of Physics, The Chinese University of Hong
Kong, Shatin, Hong Kong, China\\}
}


\begin{abstract}
We consider the spontaneous emission of a broadening molecule on
the surface of a micro-sphere in this paper. The density of states
for the micro-cavity is derived from quasi-normal models(QNM's)
expansion of the correlation functions of electromagnetic fields.
Through detailed analysis we show that only weak coupling between
a broadening atom(molecule) and the electromagnetic fields exists
in a dielectric sphere cavity whether the sphere is small or big.
From these results we find the explicit expression of the
spontaneous emission decay rate for a surfactant broadening
molecule on the surface of a micro-droplet with radius $a$, in
which only $1/a$ and $1/a^{2}$ components exhibit. Then we apply
this expression to a real experiment and obtain a consistent
result with the experiment. We also show that the real-cavity
model of local field correction is accurate, and reveal that the
local-field correction factor can be measured precisely and easily
by fluorescence experiments of surfactant molecules. Moreover, the
spontaneous decay of a surfactant molecular on droplet's surface
is sensitive to the atomic broadening, so that the fluorescence
experiment in a
micro-sphere cavity can be used to estimate the radiative broadening. \\
\end{abstract}
\maketitle
\section{Introduction}\label{sect1}
\renewcommand{\theequation}{\arabic{section}.\arabic{equation}}
It is well known that the spontaneous emission of an atom or a
molecule is not an intrinsic atomic property, but rather results
from the coupling of an atom or a molecule to the electromagnetic
environment. Purcell{\cite{Purcell}} noted first that the
spontaneous radiate rate can be enhanced if an atom is placed in a
cavity. Kleppner studied the opposite case{\cite{Kleppner}}, i.e.
inhibited spontaneous emission may happen in some conditions.
Cavity quantum electrodynamics is just to investigate the effects
of electromagnetic boundary conditions on atomic or molecular
radiative properties. A micro-sphere is easily made in experiment
not only by liquid but also by solid and acts as a
cavity{\cite{Chang,Kimble1}}. In this cavity many optical
phenomena, such as fluorescence, lasing, and many nonlinear
optical processes, have been intensively studied and a great
progress has also been achieved over the past two decades.
Recently, a great attention has been paid to the fluorescence from
spherical droplets ($\mu m$) due to a plenty of application. In
addition, the micro-droplet can be applied to biology as a
biosensor{\cite{Arnold02,Arnold03}} to detect a protein molecule.
Up to now, many studies have been made on fluorescence properties
of dye molecules in a micro-cavity
{\cite{Arnold1,Arnold2,Arnold3,Vos1}}. It has been observed that
the fluorescence decay rate shows a pronounced dependence on a
droplet radius. In particular, a surfactant molecule may be
naturally localized at the surface of liquid droplets. Their
spontaneous emission decay exhibits characteristics{\cite{Arnold2}
that were thought very much puzzling at least in the past. When
the diameter of a droplet is larger than $15\mu m$, the
spontaneous emission decay rate is a constant and much smaller
than that in bulk material. But this result cannot be explained
satisfactorily so far. We note that Arnold $et\ al$ have made some
valuable exploring and achieved important progress toward to
explaining the experimental results{\cite{Arnold4,Arnold5}}. We
also note Wu's{\cite{Wu}}
work on this topic.\\
\indent Recently, there has been much interest in the spontaneous
emission decay of an atom or a molecule embedded in material even
in dispersive and absorbing dielectric{\cite{Welsch1}}. Both
theoretical and experimental studies have been made. Since in
reality the atom embedded in dielectric is in a small region of
free space, the local field 'felt' by the atom is different from
that in the continuous medium. The decay rate is modified by local
field correction factor. Different models have been used to
calculate it. The typical models are virtual-cavity
model{\cite{V-cavity1}} and real-cavity
model{\cite{Glauber1,Tomas}}. The recent experiments have been
reported, from which the real-cavity model may be
favored{\cite{R-cavity1,R-cavity2}}. But the virtual-cavity model
is appropriate to describe interstitial gest atoms of the same
kind as the host continuents{\cite{V-cavity2}}, i.e. only one kind
of atom or molecule is present.\\
\indent It is notable that the experimental result obtained by
Barnes $et\ al${\cite{Arnold2} is normalized to the decay rate of
rhodamine B that is dissolved in the bulk glycerol. This decay
rate obviously contains affections of the medium so that the local
field correction cannot be ignored. But the surfactant molecule is
located on the air-liquid boundary, so there is no local field
correction in this case. On the other hand, in the experiment done
by Barnes $et\ al$ the emission moment orientation of surfactant
molecules is always in the tangent direction of  the surface of a
glycerol droplet, and therefore the fluorescence emission of the
dye molecule is in fact a two-dimension system, which is
considered necessary in studying its spontaneous emission decay.\\
\indent With the help of quasi-normal modes(QNM's) expansion
{\cite{Leung2}} for electric field in a sphere, we first in this
paper derive the density of states, which is a Lorentzian shape.
Then we  start from the basic equations of a two-level molecule
interacting with electromagnetic fields in full quantum theory
framework. From detailed analysis we conclude that there is no
strong coupling when a homogeneous broadening atom or molecule is
located inside or on the surface of a micro-sphere. Only weak
coupling exists in the system even if the size of a sphere is very
large. Taking these into consideration, we calculate the explicit
expression of spontaneous emission decay rate for a homogenous
broadening atom or molecule in a cavity with Lorentzian resonance
modes. Using this result and the sum rule of the density of
states, we get the very simple explicit expression of the
spontaneous emission decay rate when a molecule is on the surface
of a sphere, which indeed contains both ($1/a$) and ($1/a^{2}$)
components{\cite{Arnold3}}. Finally, we take the local-field
correction (real-cavity model) into account and draw a theoretical
curve which is in agreement with the experimental result. The
asymptotical constant of the decay rate for large diameter is
related to the local- field correction factor, which is dependent
only on the refractive index of the droplet. This result not only
verifies the accuracy of the local-field correction factor but
also shows that the fluorescence experiment of a surfactant
molecule on the surface of a sphere provides a good means of
measuring local-field correction factor. In addition, we show that
the fluorescence experiment in a micro-sphere is sensitively
dependent on the radiative broadening of atomic spectrum. This
kind of experiment can estimate the spectrum broadening. \\
\indent The present paper is organized as follows. In Section
\ref{sect2}, we derive the density of states of the
electromagnetic field on the surface of a sphere. The decay rate
of a homogeneous broadening atom or molecule in any cavity with
Lorentzian mode is given in Section \ref{sect3}. In Section
\ref{sect4}, we apply the general result to expressing explicitly
the dependence of the fluorescence decay rate on the sphere radius
and present a numerical result. Some discussions are presented in
Section \ref{sect5}.
\section {Density of states on a droplet's surface}\label{sect2}
\setcounter{equation}{0}
\subsection{Electromagnetic QNM's}
In solving the electromagnetic problems in a dielectric sphere,
the quasinormal models(QNM's) that satisfy the out-going wave
boundary condition at infinity and correspond the
morphology-dependent resonance(MDR's) are very useful physical
concept and convenient mathematical tool{\cite{Leung2}}. For a
perfect sphere, the electromagnetic fields can be expanded by
QNM's and divided into two parts, one is TE mode
\begin{eqnarray}\label{TE}
{\bf e}_{1jlm}({\bf r})=\frac{f_{1jl}}{r}{{\bf X}_{lm}}
\end{eqnarray}
and the other is TM mode
\begin{eqnarray}\label{TM}
{\bf e}_{2jlm}({\bf r})=\frac{1}{\omega_{2j}\epsilon(r)}
\nabla\times\left[\frac{f_{2jl}}{r}{\bf X}_{lm}\right]
\end{eqnarray}
where $\epsilon(r)$is the dielectric constant, ${\bf X}_{lm}$ are vector spherical harmonics
with angular momentum quantum number $l$ and magnetic quantum number $m$. $f_{\mu jl}(\mu=1,2)$
are called QNM's wave functions and satisfy the following equation($c=1$){\cite{Leung1}}
\begin{eqnarray}\label{QNMPE}
\frac{d}{dr}\beta(r)\frac{df}{dr}+\beta(r)\left[\epsilon(r)\omega^{2}
-\frac{l(l+1)}{r^{2}}\right]f(r)=0
\end{eqnarray}
where $\beta(r)=1$ for TE modes and $\beta(r)=1/\epsilon(r)$ for
TM modes. $f_{\mu jl}$ are complete and orthogonal for different
$j${\cite{Liusy}}. The sum rules
\begin{eqnarray}\label{SUM1}
\sum_{j}\frac{f_{\mu jl}(r)f_{\mu jl}(r^{'})}{\omega_{\mu j}}
=\sum_{j}\frac{f_{\mu jl}(r)f_{\mu jl}(r^{'})}{\omega_{\mu j}^{3}}=\cdots=0
\end{eqnarray}
are very useful. We will use them later.
\subsection{Density of states}
By means of QNM's, one can expand the correlation functions. It is
easy to show{\cite{Ho1}}
\begin{eqnarray}\label{DENSITY1}
{\rho}(\omega)=\frac{1}{\pi}F_{E}({\bf r,r},\omega)
\end{eqnarray}
where ${\rho}(\omega)$ is the density of states. In the free space, it is well known that
\begin{eqnarray}\label{DENSITY2}
{\rho}_{0}(\omega)=\frac{\omega^{3}}{\pi^{2}c^{3}}
\end{eqnarray}
$F_{E}({\bf r,r},\omega)$ is the
sum of vacuum fluctuation of the electric field for all components
\begin{eqnarray}\label{DENSITY3}
{\rho}(\omega)&=&\frac{1}{\pi}\sum_{i}F_{Eii}({\bf r,r},\omega)\nonumber\\
&=&\omega^{2}{\rm Im}\sum_{lj}\frac{{\bf e}_{\mu jlm}({\bf r})
\cdot{\bf e}_{\mu jlm}^{\dagger}({\bf r})}
{\omega_{jl}(\omega-\omega_{jl})}
\end{eqnarray}
where ${\bf e}_{\mu jlm}^{\dagger}({\bf r})$ is the conjugate
vector of ${\bf e}_{\mu jlm}({\bf r})$, in which only ${\bf
X}_{lm}$ is replaced by ${\bf X}_{lm}^{*}$. QNM's eigenfunctions
for a fixed $l$ are given by
\begin{eqnarray}\label{FJ}
f_{j}(r)=\left\{
\begin{array}{cc}
C_{j}rj_{l}(n_{0}\omega_{j}r)&\mbox{}\hspace{10mm}0<r<a\\
B_{j}rh_{l}^{(1)}(\omega_{j}r)&\mbox{}\hspace{16mm}r>a
\end{array}
\right.
\end{eqnarray}
The coefficients $C_{j}$ are
\begin{eqnarray}\label{COE1}
C_{j}^{2}=\left\{
\begin{array}{ll}
\left[a^{3}(n_{0}^{2}-1)j_{l}(n_{0}\omega_{j}a)/2\right]^{-1}
            &\mbox{}\hspace{4mm}{\rm TE\ case}\\
\left[a^{3}(n_{0}^{2}-1)S/2\right]^{-1}&\mbox{}\hspace{4mm}{\rm
TM\ case}
\end{array}
\right.
\end{eqnarray}
where
\begin{eqnarray}\label{COE2}
S=\frac{1}{n_{0}^{2}}
\left\{
    \left[\frac{j_{l}^{'}(n_{0}\omega_{j}a)}{j_{l}(n_{0}\omega_{j}a)}
    +\frac{1}{n_{0}\omega_{j}a}\right]^{2}
+\frac{l(l+1)}{(\omega_{j}a)^{2}}
\right\}
\end{eqnarray}
and
\begin{eqnarray}\label{COE3}
B_{j}=C_{j}j_{l}(n_{0}\omega_{j}a)/h_{j}^{(1)}(\omega_{j}a)
\end{eqnarray}
In the above equations, the resonance frequency $\omega_{j}$ is complex
\begin{eqnarray}\label{RESFRE1}
\omega_{j}=\omega_{j0}+i\omega_{jI}
\end{eqnarray}
The real part is the resonance frequency of MDR's and the
imaginary part is half of the full width at half maximum(HFWHM) of
these resonances. We can rewrite $\omega_{j}$ as the common form
\begin{eqnarray}\label{RESFRE2}
\omega_{j}=\omega_{j0}+i\frac{\gamma_{lj}}{2}
\end{eqnarray}
The electric field for TE mode is only along the tangent of
droplet's surface. Using  (\ref{DENSITY3}) and sum rules
(\ref{SUM1}), we can get the density of states
\begin{eqnarray}\label{DENSITY4}
{\rho}(\omega)&=&\sum_{lj}\frac{2l+1}{4\pi^{2}}\frac{2\omega}{a^{3}(n_{0}^{2}-1)}
\frac{1}{\omega_{jI}}\frac{(\omega_{jI})^{2}}{(\omega-\omega_{j0})^{2}
+(\omega_{jI})^{2}}\nonumber\\
&=&\sum_{lj}{\rho}_{0}(\omega)\frac{2l+1}{(n_{0}^{2}-1)x^{2}}\frac{1}{\gamma_{lj}^{x}}
\frac{(\gamma_{lj}/2)^{2}}{(\omega-\omega_{lj})^{2}+(\gamma_{lj}/2)^{2}}\nonumber\\
&\approx&{\rho}_{0}(\omega)\sum_{lj}\frac{2l+1}{(n_{0}^{2}-1)x_{lj}^{2}}\frac{1}{\gamma_{lj}^{x}}
\frac{(\gamma_{lj}^{x}/2)^{2}}{(x-x_{lj})^{2}+(\gamma_{lj}^{x}/2)^{2}}.\nonumber\\
\end{eqnarray}
where $x$,\ $x_{lj}$ and $\gamma_{lj}^{x}$ are the dimension-free variables(restore $c$)
\begin{eqnarray}\label{RESFRE3}
x&=&\frac{\omega}{c}a\nonumber\\
x_{lj}&=&\frac{\omega_{j0}}{c}a\nonumber\\
\gamma_{lj}^{x}&=&\frac{\gamma_{j0}}{c}a
\end{eqnarray}
$x_{lj}$ and $\gamma_{lj}^{x}$ are independent of sphere radius $a$ and $\gamma_{lj}^{x}$
augment along with $j$ for a fixed $l$ and are given asymptotically by{\cite{Leung1}}
\begin{eqnarray}\label{RESWID}
\gamma_{l\infty}^{x}=\frac{1}{n_{0}}ln\frac{n_{0}+1}{n_{0}-1}
\end{eqnarray}
So the density of states on the surface of a sphere is the
standard Lorentzian distribution. For a given $l$ there are
infinity resonance modes, but only several resonance modes ahead
($j=1,2,3\dots$) are important because these resonance widths are
narrower and the resonance points $x_{lj}$ are smaller.
\section{Basic equations and their solutions}\label{sect3}
\setcounter{equation}{0}
Ten years ago{\cite{Leung3}}, Lai $et\
al$ studied the spontaneous decay rate of a two-level atom in a
cavity with Lorentzian modes and discussed the conditions of weak
coupling, strong coupling and intermediate coupling. In this
section, we still use the same method and investigate the
radiative properties of a homogeneous broadening molecule. We
consider a system that is composed of an atom, electromagnetic
fields(cavity) and their interaction. The atom, which has two
energy levels, a lower level $a$ and upper level $b$, is located
at position ${\bf r_{0}}$ in a cavity. The interaction Hamiltonian
between the atom and the electric field is
\boldmath
\begin{eqnarray}\label{HAMIL}
\mbox{\unboldmath{V}}={\mu\cdot E(r_{0})}
\end{eqnarray}
where ${\bf\mu}$\unboldmath\  is the electric dipole operator and $\bf E$ is the electric field.
We assume that an excited atom is initial at upper level
$|b\rangle$ and $C(t)$ denotes it's amplitude. The other related
state of the system is described by $|as\rangle$ with amplitude
$D_{s}(t)$, in which the atom is at the lower atomic state
$|a\rangle$ and one photon is in mode $s$. In interaction
representation and under the rotating wave approximation, one can
obtain usual Wigner-Weisskopf equation{\cite{Leung3, Heitler}}
\begin{eqnarray}\label{WWE}
i\hbar\frac{dC(t)}{dt}=\sum_{s}V_{s}^{*}D_{s}e^{i(\Omega-\omega_{s})t}\\
i\hbar\frac{dD_{s}(t)}{dt}=V_{s}C(t)e^{-i(\Omega-\omega_{s})t}
\end{eqnarray}
where $\hbar\Omega=E_{b}-E_{a}$ and $V_{s}=\langle as|V|b\rangle$.
After some algebraic calculations, it is easy to get
\begin{eqnarray}\label{C1}
\frac{dC(t)}{dt}=-\frac{2\pi}{\hbar}\frac{M}{3}\int\limits_{0}^{\infty}d\omega
{\rho}(\omega)\int_{0}^{t}d\tau C(\tau)e^{i(\Omega-\omega)(t-\tau)}
\end{eqnarray}
where
\begin{eqnarray}
M=\langle a|\mu_{i}|b\rangle \langle b|\mu_{i}|a\rangle
\end{eqnarray}
and $\rho(\omega)$ is the density of states at position ${\bf
r_{0}}$ in the cavity. The factor $3$ appears in (\ref{C1})
because we have assumed that the atomic dipole matrix element is
isotopic and has equal probability in any
direction in 3-dimension space.\\
\indent For complex molecules or homogeneous broadening atoms, the
fluorescence spectrum is band-type, so one must consider a large
number of lower levels $a_{1}, a_{2}, \cdots$ forming a continuum.
Then the amplitude of the probability that the atom is in up state
should satisfy
\begin{eqnarray}\label{C2}
\frac{dC(t)}{dt}&=&-\frac{2\pi}{\hbar}\frac{1}{m}\int\limits_{0}^{+\infty}d\omega
{\rho}(\omega)\nonumber\\
&*&\int\limits_{0}^{+\infty} d\Omega\int_{0}^{t}d\tau C(\tau)M(\Omega)
e^{i(\Omega-\omega)(t-\tau)}
\end{eqnarray}
where $m$ is the number of freedom degree of the atomic dipole in
a cavity and $M(\Omega)$ is atomic dipole matrix element per unit
transition frequency. In practical cases, the diople moment of an
atom or a molecule is not free in every direction, so we must
introduce this parameter $m$. Assuming that the line shape of
atomic braodening is Lorentzian distribution
\begin{eqnarray}\label{lineshape}
M(\Omega)=M_{0}\frac{\Gamma_{h}}{2\pi}\frac{1}
{(\Omega-\Omega_{0})^{2}+(\frac{\Gamma_{h}}{2})^{2}}
\end{eqnarray}
we now derive that the strong coupling condition when a broadening
atom is in a sphere cavity. Considering the ideal case, i.e. the
maximum coupling $\Omega_{0}=\omega_{0}$, one can regard the
cavity as a single mode cavity
\begin{eqnarray}\label{DENSITY5}
\rho(\omega)=\rho_{0}K\frac{(\frac{\gamma}{2})^{2}}{(\omega-\omega_{0})^{2}+(\frac{\gamma}{2})^{2}}
\end{eqnarray}
where
\begin{eqnarray}\label{K1}
K=\frac{2l+1}{n_{0}^{2}-1}\frac{1}{x_{0}^{2}}\frac{1}{\gamma^{x}}
\end{eqnarray}
Then $C(t)$ satisfies
\begin{eqnarray}\label{C3}
\frac{dC(t)}{dt}=\int_{0}^{t}d\tau S(t-\tau)C(\tau)
\end{eqnarray}
where the kernel $S$ is
\begin{eqnarray}\label{S}
S(t-\tau)=-\frac{K\gamma}{4\tau_{0}}e^{-\frac{1}{2}(\Gamma_{h}+\gamma)(t-\tau)}
\end{eqnarray}
$\tau_{0}$ is the spontaneous decay lifetime in vacuum
\begin{eqnarray}\label{TAU0}
\tau_{0}=\frac{4M_{0}\Omega_{0}^{3}}{3\hbar c^{3}}
\end{eqnarray}
It is easy to find
\begin{eqnarray}\label{C4}
\frac{d^{2}C(t)}{dt^{2}}+\frac{1}{2}(\Gamma_{h}+\gamma)\frac{dC(t)}{dt}
+\frac{K\gamma}{4\tau_{0}}C(t)=0
\end{eqnarray}
The strong coupling should satisfy the following condition{\cite{Leung3}}
\begin{eqnarray}\label{STRC1}
\frac{K\gamma}{\tau_{0}}\gg\left(\frac{\Gamma_{h}+\gamma}{2}\right)^2
\approx\frac{\Gamma_{h}^{2}}{4}
\end{eqnarray}
\\
Given $x_{0}=\Omega_{0}a/c$, when $l\approx n_{0}x_{0}$(the least
leaky mode){\cite{Leung1,Leung4}},
the enhanced factor $K$ in (3.9) will be maximum
\begin{eqnarray}\label{K2}
K_{max}=\frac{2n_{0}}{n_{0}^{2}-1}\frac{1}{x_{0}}\frac{1}{\gamma^{x}}
\end{eqnarray}
the strong coupling condition reduces
\begin{eqnarray}\label{STRC2}
\left(\frac{\Gamma_{h}}{c}\right)^{2}\ll\frac{8n_{0}}{n_{0}^{2}-1}\frac{1}{x_{0}a\tau_{0}c}
\end{eqnarray}
In visible light domain, $\tau_{0}\sim 10^{-9}s$, $a\sim 10\mu m$ and $x_{0}\sim 100$,
the above inequality becomes
\begin{eqnarray}\label{STRC3}
\left(\frac{\Gamma_{h}}{c}\right)^{2}\ll 10(cm^{-1})
\end{eqnarray}
So the strong coupling will be exhibited only when
$\Gamma_{h}<1cm^{-1}$, which is very difficult to be met in
experiment because the typical value of the homogeneous broadening
is $100cm^{-1}$. If the atomic broadening is in the order of
$\Gamma_{h}\sim 10cm^{-1}$,
the interaction between an atom or a molecule and a spherical cavity must be weak coupling.\\
\indent For Rydberg's atom, $\tau_{0}\sim 10^{-2}s$, $a\sim 1cm$
and $x_{0}\sim 100$, the strong coupling condition is changed into
\begin{eqnarray}\label{STRC4}
\left(\frac{\Gamma_{h}}{c}\right)^{2}\ll 10^{-8}(cm^{-1})
\end{eqnarray}
Which is also very difficult to be satisfied. So we can conclude
that there is no strong
coupling between a broadening atom and microsphere cavity modes.\\
\indent Under the weak coupling case, it is significant only for
$t-\tau\rightarrow 0$ in the integration of (\ref{C2}), so that it
becomes Markovian
\begin{eqnarray}\label{c5}
\frac{dC(t)}{dt}\approx -\left[\frac{\gamma}{2}+i\delta\right]C(t)
\end{eqnarray}
and
\begin{eqnarray}\label{gamma1}
\frac{\gamma}{2}+i\delta=\frac{2\pi}{\hbar}\frac{1}{m}\int d\omega\int d\Omega
\frac{{\rho}(\omega) M(\Omega)}{i(\omega-\Omega-i\epsilon)}
\end{eqnarray}
We assume that $\rho(\omega)$ is still Lorentzian
\begin{eqnarray}\label{density6}
\rho(\omega)=\rho_{0}\sum_{j}K_{j}\frac{(\frac{\gamma_{j}}{2})^{2}}
{(\omega-\omega_{j})^{2}+(\frac{\gamma_{j}}{2})^{2}}
\end{eqnarray}
Substituting (\ref{lineshape}) and above equation into (\ref{gamma1}), we have
\begin{eqnarray}\label{gamma2}
\frac{\gamma}{2}&=&\rho_{0}\frac{2\pi^{2}}{\hbar}\frac{M_{0}\Gamma_{h}}{2m\pi}
\sum_{j}K_{j}\nonumber\\
&&*\int\limits_{0}^{+\infty}d\omega
\frac{(\frac{\gamma_{j}}{2})^{2}}{(\omega-\omega_{j})^{2}+(\frac{\gamma_{j}}{2})^{2}}
\frac{1}{(\omega-\Omega_{0})^{2}+(\frac{\Gamma_{h}}{2})^{2}}\nonumber\\
&=&\frac{1}{\tau_{0}}\frac{3}{8m}\sum_{j}K_{j}\gamma_{j}
\frac{\Gamma_{h}+\gamma_{j}}{(\Omega_{0}-\omega_{j})^{2}+
(\frac{\Gamma_{h}+\gamma_{j}}{2})^{2}}\nonumber\\
&=&\frac{1}{\tau_{0}}\frac{3}{2m}\sum_{j}K_{j}\gamma_{j}\frac{\Gamma_{h}+\gamma_{j}}
{(2\Delta_{j})^{2}+(\Gamma_{h}+\gamma_{j})^{2}}
\end{eqnarray}
where $\Delta_{j}=\Omega_{0}-\omega_{j}$ and we have used the integration formula
\begin{eqnarray}\label{integ3}
\int\limits_{0}^{+\infty}d\omega
\frac{1}{(\omega-\omega_{j})^{2}+(\frac{\gamma_{j}}{2})^{2}}
\frac{1}{(\omega-\Omega_{0})^{2}+(\frac{\Gamma_{h}}{2})^{2}}\nonumber\\
=\frac{2\pi}{\gamma_{j}\Gamma_{h}}\frac{\Gamma_{h}+\gamma_{j}}
{(\Omega_{0}-\omega_{j})^{2}+(\frac{\Gamma_{h}+\gamma_{j}}{2})^{2}}
\end{eqnarray}
in which we have assumed $\gamma_{j},\Gamma_{h}\ll \omega_{j},\Omega$, so that the
integral can be evaluated by integrating over the whole real line.\\
\indent The spontaneous emission decay rate of a homogeneous
broadening molecule in the cavity is finally written in the
following form
\begin{eqnarray}\label{gamma3}
\frac{\gamma}{\gamma_{0}^{vac}}=\frac{3}{m}\sum_{j}K_{j}\gamma_{j}\frac{\Gamma_{h}+\gamma_{j}}
{(2\Delta_{j})^{2}+(\Gamma_{h}+\gamma_{j})^{2}}
\end{eqnarray}
where $\gamma_{0}^{vac}=1/{\tau_{0}}$ is the decay rate of the
molecule in vacuum. For resonance case $\Delta_{j}=0$ and if the
cavity mode spacing is much larger than $\Gamma_{h}$, only one
cavity mode is important and then the decay rate can be
approximately expressed as
\begin{eqnarray}\label{gamma4}
\frac{\gamma}{\gamma_{0}^{vac}}\approx\frac{3}{m}\frac{K\gamma}{(\Gamma_{h}+\gamma)}
\end{eqnarray}
Letting $\delta_{c}$ be the cavity modes separation and using the
sum rule {\cite{Ching1,Yokoyama1}}
\begin{eqnarray}\label{sum2}
K\gamma\approx \delta_{c}
\end{eqnarray}
we can obtain the two limiting cases for m=3
\begin{eqnarray}\label{gamma5}
\frac{\gamma}{\gamma_{0}^{vac}}=
\left\{
\begin{array}{ll}
    K \mbox{}\hspace{20mm}\Gamma_{h}\ll\gamma\\
    \delta_{c}/\Gamma_{h}\mbox{}\hspace{14mm}\Gamma_{h}\gg\gamma
\end{array}
\right.
\end{eqnarray}
which are just the common results of weak
coupling{\cite{Yokoyama1}}. The above equations are held only when
the atom is resonated in a cavity and the cavity mode spacing is
much larger than the resonance width of a cavity mode. But
(\ref{gamma3}) is a accurate expression of spontaneous emission
decay rate for a broadening molecule in a cavity under the
weak coupling condition.\\
\indent Generally speaking, the enhanced factor $K_{j}$ satisfies
$K_{j}=K_{j}^{0}/\gamma_{j}$, so that $K_{j}\gamma_{j}=K_{j}^{0}$
is independent of $\gamma_{j}$ and then
\begin{eqnarray}\label{gamma6}
\frac{\gamma}{\gamma_{0}^{vac}}=\frac{3}{m}\sum_{j}\frac{K_{j}^{0}}{\gamma_{j}^{'}}
\frac{(\gamma_{j}^{'}/2)^{2}}
{(\Omega_{0}-\omega_{j})^{2}+(\gamma_{j}^{'}/2)^{2}}
\end{eqnarray}
where $\gamma_{j}^{'}=\gamma_{j}+\Gamma_{h}$. This means that the
spontaneous emission decay rate is formally proportional to the
density of fields states but the resonance width is replaced by
the sum of a cavity mode width and the broadening of a molecule's
energy level, which may be very useful to the analysis of the
decay rate of a homogeneous broadening molecule in practice.\\
\indent In the following section, we shall concentrate on
spherical cavities and explain the experimental result.
\section {Spherical cavities cases}\label{sect4}
\setcounter{equation}{0}
We now apply the general results of the former section to
a homogeneous broadening molecule in a spherical cavity. For a dielectric sphere,
the expression (\ref{integ3}) of the decay rate is given by the following form
\begin{eqnarray}\label{gamma41}
\frac{\gamma}{\gamma_{0}^{vac}}=\frac{3}{m}\sum_{li}K_{li}\gamma_{li}^{x}
\frac{\Gamma_{h}^{x}+\gamma_{li}^{x}}
{(2\Delta x_{li})^{2}+(\Gamma_{h}^{x}+\gamma_{li}^{x})^{2}}
\end{eqnarray}
In the practical experiment, the result is
normalized to the decay rate of rhodamine B in bulk glycerol. But $\gamma_{0}^{vac}$
is the decay rate of a broadening molecule in vacuum. In order to compare the
theoretical result with the experimental data, both sides of the above equation
should be divided by the factor
$n_{0}\xi_{lc}$ and then
\begin{eqnarray}\label{gamma42}
\frac{\gamma}{\gamma_{0}}&=&\frac{1}{n_{0}\xi_{lc}}\frac{3}{m}
\sum_{li}K_{li}\gamma_{li}^{x}\frac{\Gamma_{h}^{x}
+\gamma_{li}^{x}}
{(2\Delta x_{li})^{2}+(\Gamma_{h}^{x}+\gamma_{li}^{x})^{2}}\nonumber\\
&=& \left.
        \frac{1}{n_{0}\xi_{lc}}\frac{3}{m}\frac{\rho_{c}(x_{0})}{\rho_{0}}
    \right|_{\gamma_{li}^{x}\rightarrow\gamma_{li}^{x}+\Gamma_{h}^{x}}
\end{eqnarray}
where $\xi_{lc}$ is the local-field correction factor and according to real-cavity model
{\cite{Welsch1,Glauber1}} it reads
\begin{eqnarray}\label{rcf}
\xi_{RC}=\left(\frac{3n_{0}^{2}}{2n_{0}^{2}+1}\right)^{2}
\end{eqnarray}
Because the surfactant molecule is on the surface of a sphere, the
field 'felt' by a surfactant molecule is the same as the
macro-field on the surface of a sphere, so there is no local-field
correction for the decay rate of a surfactant molecule in this
case. From (\ref{gamma42}), it is easy to see that the peak values
of the spontaneous emission decay rate are all at the positions of
MDR's, but the widths are much larger than those of MDR's. Dye
molecules are of multi-atom molecules so that the fluorescence
spectrum is very wide. For octadecyl rhodamine B used by Barns
$et\ al${\cite{Arnold2}} in their experiment, the center of its
fluorescence spectrum is at $\lambda_{0}=560nm$ with the band
width of $\Delta\lambda=60nm$. Within the band width, we note that
the interresonance separation is about
$\delta_{c}^{x}=0.7${\cite{Leung4}} for a spherical cavity with
$n_{0}=1.47$. Therefore 3 or more transitions are resonance
transitions, in which the transition probabilities are much larger
than those non-resonance transitions. So the spontaneous emission
decay rate is determined mainly by these resonance transitions, in
which the corresponding transition probabilities are increased
with the decrease of the transition wave length. It is a good
approximation to choose the decay rate of a dye molecule whose the
resonance transition is near $\lambda_{0}$ as the average value of
all resonance transitions. On the other hand, the directions of a
surfactant molecular dipole moment are always along the tangency
of the sphere so that $m=2$ and then TE mode is much more
significant than TM mode for spontaneous emission. In
(\ref{gamma42}), because
$\delta_{c}^{x}\gg\gamma_{li}^{x}+\Gamma_{h}^{x}$, the least leaky
cavity mode ($i=1$ and $l\approx n_{0}x_{0}$) is the most
important for given $x_{0}=x_{li}$, the other parts can be
regarded as 'background'
\begin{eqnarray}\label{density7}
\left. \frac{\rho_{c}(x_{0})}{\rho_{0}}
\right|_{\gamma_{li}^{x}\rightarrow\gamma_{li}^{x}+\Gamma_{h}^{x}}
=\gamma_{b}+\frac{K_{l1}\gamma_{l1}^{x}(\Gamma_{h}^{x}+\gamma_{l1}^{x})}
{(2\Delta x_{l1})^{2}+(\Gamma_{h}^{x}+\gamma_{l1}^{x})^{2}}
\end{eqnarray}
where the second term in the above equation is the contribution of
the least leaky TE resonance mode that is near to $x_{0}$, while
the first term $\gamma_{b}$ is called 'background', in which all
TE modes and the tangent parts of TM modes are included except the
least leaky TE resonance mode $x_{l1}$.\\
On the surface of a sphere, the density of states is redistributed
and the sum rule {\cite{Ching1}}
\begin{eqnarray}\label{sum3}
\int\limits_{x_{0}-\delta_{c}^{x}/2}^{x_{0}+\delta_{c}^{x}/2}dx_{0}
\frac{\rho_{c}(x_{0})}{\rho_{0}}=\delta_{c}^{x}
\end{eqnarray}
is still valid, because
$\delta_{c}^{x}\gg\gamma_{li}^{x}+\Gamma_{h}^{x}$. Here
$\delta_{c}^{x}$ is the cavity mode spacing . Substituting
(\ref{density7}) into the above equation, one can easily show
\begin{eqnarray}\label{gammab}
\gamma_{b}=1-\frac{\pi}{2}\frac{K_{l1}\gamma_{l1}^{x}}{\delta_{c}^{x}}
\end{eqnarray}
Using (\ref{DENSITY4}), we have
\begin{eqnarray}\label{sum4}
K_{l1}\gamma_{l1}^{x}&=&\frac{2l+1}{n_0^{2}-1}\frac{1}{x_{l1}^{2}}\nonumber\\
&\approx&\frac{2n_{0}}{n_{0}^{2}-1}\frac{1}{x_{0}}
\end{eqnarray}
and then
\begin{eqnarray}\label{gamma43}
\frac{\gamma}{\gamma_{0}}=\frac{1}{n_{0}\xi_{lc}}\frac{3}{2}
\left[
    1+\frac{2n_{0}}{n_{0}^{2}-1}\frac{1}{x_{0}}
        \left(
            \frac{1}{\Gamma_{h}^{x}}-\frac{\pi/2}{\delta_{c}^{x}}
        \right)
\right]
\end{eqnarray}
where we have omitted $\gamma_{l1}^{x}$ and let $m=2$. If we let
\begin{eqnarray}
x_{0}&=&\frac{2\pi}{\lambda_{0}}a=\alpha a\\
\Gamma_{h}^{x}&=&\frac{\Gamma_{h}}{c}a=\frac{2\pi\Delta\nu_{s}}{c}a=\beta a
\end{eqnarray}
where $a$ will be in the unit of $\mu m$ later. (\ref{gamma43}) is
rewritten as the following form
\begin{eqnarray}\label{gamma44}
\frac{\gamma}{\gamma_{0}}=\frac{1}{n_{0}\xi_{lc}}\frac{3}{2}
\left[
    1+\frac{2n_{0}}{n_{0}^{2}-1}\frac{1}{\alpha}
        \left(
            \frac{1}{\beta a^{2}}-\frac{\pi/2}{\delta_{c}^{x}}\frac{1}{a}
        \right)
\right]
\end{eqnarray}
In the above equation, there are $1/a$ and $1/a^{2}$ components,
the result of which is in agreement with
what was obtained by Arnold{\cite{Arnold4}}.\\
\indent Arnold estimates from their experiment that the
homogeneous broadening of rhodamine B in glycerol is
$\Delta\nu=100cm^{-1}${\cite{Arnold2,Arnold4}}. Because the
surfactant molecule is on the surface of droplets of glycerol, it
is reasonable to think that the homogeneous broadening of a
surfactant molecule is only half of that in bulk glycerol. If we
take $\Delta\nu_{s}=50cm^{-1}$, $\beta=0.0314$. The other
parameter $\alpha=11.2$ for $\lambda_{0}=560nm$. The refractive
index of glycerol is $n_{0}=1.47$. By means of these parameters,
we can easily get the decay rate for any size spherical cavity.
Fig.1 is the numerical result. In Fig.1, the three curves
correspond to three different local-field correction factors. The
experimental data show that the local-field correction is slightly
less ($90\%-95\%$) than that of real-cavity model. This difference
maybe coming from the fact that the size of the dye molecule is
much bigger than single atom molecule, so that the local-field
correction factor of the real-cavity model is just a good
approximation. The affections of the molecular shape should be
taken into account for this case.\\
\begin{figure}
\includegraphics{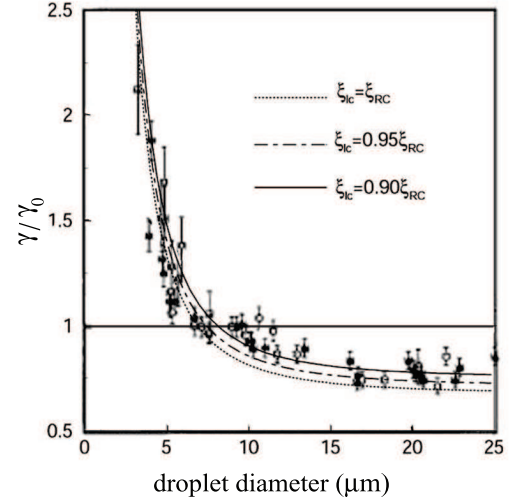}
\caption{The spontaneous emission decay rate of octadecyl
rhodamine B on the surface of a glycerol droplet as a function of
droplet size. Three curves correspond to the three different
local-field correction factors. We take $\Delta\nu_{s}=50cm^{-1}$
in Eq.(4.10). The experimental data are from
Ref.{\cite{Arnold2}.}}
\end{figure}
\indent
From (\ref{gamma44}), the local-field correction factor is connected easily with the experimental result
\begin{eqnarray}\label{rcf2}
\xi_{lc}=\frac{3}{2n_{0}g}
\end{eqnarray}
where
\begin{eqnarray}\label{g}
g=\lim_{a\rightarrow\infty}\frac{\gamma}{\gamma_{0}}
\end{eqnarray}
If one measures the decay rate of surfactant molecules on the
surface of a larger sphere, the local-field correction factor is
then given by (\ref{rcf2}). This reveals that the fluorescence
experiment of surfactant molecules is a good approach to measure
precisely and conveniently the local-field correction factor,
which was not recognized and is out of what we have predicted before.\\
\indent In (\ref{gamma43}), apart from $\Gamma_{h}$, other
parameters are determined only by the refractive index $n_{0}$, in
which $\delta_{c}^{x}$ is derived from Mie's
theory{\cite{Leung4}}. What we have chosen in Fig.1 for
$\Gamma_{h}$ is based on the above analysis. In fact, the decay
rate is very sensitive to $\Gamma_{h}$ because $\beta$ is in the
component of $1/a^{2}$ and then the room where $\Gamma_{h}$ may be
chosen in this experiment is very small. The numerical
calculations show that $\Gamma_{h}$ in Fig.1 is really the best.
So this experiment is also a good method to measure or estimate at
least the homogeneous broadening of a molecule.
\section{Conclusions and discussions}\label{sect5}
\setcounter{equation}{0}
In the present paper we have considered  the spontaneous emission
decay rate of a broadening atom or molecule in a cavity. For a
broadening molecule, it is very difficult to exhibit the strong
coupling between molecule and fields in a cavity to happened. We
have proved by detailed analysis that there is no strong coupling
in a dielectric sphere cavity. Under the weak coupling condition,
we obtained the exact expression of spontaneous emission decay
rate for a broadening molecule in any cavity with Lorentzian mode
distribution and discussed the two common limiting cases. Simply
speaking, the decay rate of spontaneous emission for a homogeneous
broadening molecule equals the density of states, in which the
widths of the cavity resonance modes are replaced by the sum of
cavity mode width and the broadening of a atomic energy level.
Applying this general formula to a spherical cavity, we have a
simple analytical expression that shows explicitly the dependence
of the fluorescence decay rate on the spherical radius. When we
explain the experimental result\cite{Arnold2}, two points are
important: one is that the freedom degree of the surfactant
molecule transition dipole moment is $2$ and the other is the
local-field correction factor. Moreover, QNM's expansion of the
fields is very convenient to obtain the density of states for a
spherical cavity. The sum rule of the density of states plays a
very important role in simplifying the result and in numerical
calculation. Finally, it is noted that new significance of the
fluorescence experiment on the surface of droplets is revealed in
our present work. This kind of experiment is a very good method to
measure precisely the local-field correction factor.\\
\indent We gratefully acknowledge helpful discussions with Prof.
S. Arnold. Y-J. Xia and G-C. Guo were supported by the National
Fundamental Research Program (2001CB309300), National Natural
Science Foundation of China, and the Innovation funds from
the Chinese Academy of Sciences. \\

\end {document}